# Hadoop-Oriented SVM-LRU (H-SVM-LRU): An Intelligent Cache Replacement Algorithm to Improve MapReduce Performance

Rana Ghazali [a], Sahar Adabi [a], Ali Rezaee[b], Douglas G.Down [c], Ali Movaghar [d]

R_ghazali@iau-tnb.ac.ir, Sahar_adabi@iau-tnb.ac.ir, Alirezaee@srbiau.ac.ir, Downd@mcmaster.ca, Movaghar@sharif.edu

a: Department of Computer Engineering, North Tehran Branch, Islamic Azad University, Tehran, Iran
b: Department of Computer Engineering, Science and Research Branch, Islamic Azad University, Tehran, Iran
c: Department of Computing and Software, McMaster University, 1280 Main St W, Hamilton, ON, Canada
d: Department of Computer Engineering, Sharif University of Technology, Azadi Ave, Tehran, Iran

**Abstract**

Modern applications can generate a large amount of data from different sources with high velocity, a combination that is difficult to store and process via traditional tools. Hadoop is one framework that is used for the parallel processing of a large amount of data in a distributed environment, however, various challenges can lead to poor performance. Two particular issues that can limit performance are the high access time for I/O operations and the recomputation of intermediate data. The combination of these two issues can result in resource wastage. In recent years, there have been attempts to overcome these problems by using caching mechanisms. Due to cache space limitations, it is crucial to use this space efficiently and avoid cache pollution (the cache contains data that is not used in the future). We propose Hadoop-oriented SVM-LRU (H-SVM-LRU) to improve Hadoop performance. For this purpose, we use an intelligent cache replacement algorithm, SVM-LRU, that combines the well-known LRU mechanism with a machine learning algorithm, SVM, to classify cached data into two groups based on their future usage. Experimental results show a significant decrease in execution time as a result of an increased cache hit ratio, leading to a positive impact on Hadoop performance.

## 1 Introduction

Hadoop [1] is an open-source framework for the storage and parallel processing of large datasets. Two major components of Hadoop are HDFS (Hadoop Distributed File System) and MapReduce. HDFS is a distributed file system with a master/slave architecture. Input data are split into data blocks of identical size, distributing multiple copies (according to a replication factor) of each block to different machines to provide fault tolerance. MapReduce is a method for the parallel processing of a large amount of data on a cluster of machines in a distributed environment. MapReduce consists of three phases: Map, Shuffle and Reduce. First of all, Map tasks convert input data into <key, value> pairs (referred to as intermediate data) then these intermediate data are sorted, shuffled, and provided as input for the Reduce tasks. Finally, Reduce tasks merge values with identical keys to generate final results.

Hadoop is a popular platform for analyzing various data types (structured, semi-structured, and unstructured data) and a number of Big Data tools are designed based on this platform. Hadoop has some advantages over Relational DataBase Management Systems (RDBMS) such as flexibility, high throughput, low cost, and concurrent processing. Also, when we compare Hadoop with other alternatives for Big Data processing such as Spark, we observe some plus points like security, scalability, and high fault tolerance. However, Hadoop has some challenges that can lead to poor performance:
1. HDFS is based on a hard disk drives (HDD) system. The high access times for I/O operations can have a significant impact on the overall execution time [2].
2. The shuffle phase in MapReduce is a time-consuming operation: as much as 33% of the overall execution time is spent on this phase [3].
3. A large amount of intermediate data is thrown away after processing; this requires recomputation if reuse is required [4].
4. The MapReduce programming model is not well-suited for iterative programs. A large amount of intermediate data may be unchanged from one iteration to the next. The lack of a mechanism to identify duplicate computations means that the data must be re-loaded and re-computed at each iteration leading to wasted I/O, network bandwidth, and CPU resources. The second problem is related to identifying termination conditions via a fixed point corresponding to the application's output not changing for successive iterations. This itself needs an extra MapReduce job for each iteration, degrading performance [5].

In recent years, many researchers have proposed the use of caching mechanisms to address these challenges [2], [3], [4], [5], [6], [7]. By use of a caching mechanism, data is prefetched into cache memory in order to reduce overall execution time. A caching mechanism consists of two phases: the placement phase and the delivery phase. The placement phase determines how to place data into the cache memory according to some measure of data popularity. A limited cache size creates the need for a replacement policy that determines how to remove content from the cache if new content is to be added (when the cache capacity is reached). There are a number of different replacement algorithms that have been proposed. The second phase is the delivery phase, which retrieves data from cache memory according to user demands. Network congestion may result if user demands are sufficiently high. While a caching strategy can have a positive impact on Hadoop's performance, to maximize this impact the limited cache space must be efficiently used, in particular, minimizing cache pollution is a key goal.

While a decision if data should be cached must be made, the more important decision is which data should be replaced if new data is to be added to a full cache. Therefore, cache replacement is the core of caching. In this paper, we propose a Hadoop-oriented SVM-LRU (H-SVM-LRU) approach that applies the intelligent cache replacement algorithm SVM-LRU to optimize the use of cache space and avoid cache pollution. A machine learning component classifies cached data into two groups (reused in the future or not) to recognize which data should remain in the cache and which data should be replaced. The goal is to use limited cache space efficiently resulting in a positive impact on the cache-hit ratio. As a result, this method can be appropriate for iterative programs by both reducing data access time from disk and avoiding recomputation through accessing more intermediate data from the cache.

Our contributions in this paper are:
- We provide an overview of intelligent cache replacement methods used in web proxy caches.
- Different cache replacement strategies are investigated in the Hadoop environment, and we discuss their advantages and disadvantages.
- We introduce the intelligent caching mechanism, H-SVM-LRU, in the Hadoop environment, which uses SVM for classifying data into two groups: reused in the future or not.
- We evaluate H-SVM-LRU's performance (hit ratio) and compare it with LRU.

- We carry out experiments to investigate the impact of this algorithm on Hadoop's execution time performance.

The rest of the paper is organized as follows: Section 2 defines the problem and our solution approach. We discuss existing caching replacement strategies for Hadoop and intelligent caching methods for web proxy caches in Section 3. We then describe the H-SVM-LRU framework and present our H-SVM-LRU algorithm for the Hadoop environment in Section 4. Next, we explain the details of the H-SVM-LRU implementation in Section 5. We evaluate the performance of H-SVM-LRU via different experiments in Section 6. Finally, Section 7 contains the conclusions and suggestion for future work.

## 2 Problem definition

Reducing job execution time is an effective factor for improving Hadoop performance. The execution time is composed of two components: I/O operation time and processing time. Since input data are stored in an HDD-based system, HDFS, access time for I/O operations can be high, adversely affecting execution time. One approach to tackle this problem is to use a caching mechanism to store required data in the cache, however, the cache has limited space which must be effectively managed.

Hadoop 2.3.0 has been released with support the in-memory caching to mitigate the cost of I/O operations and increase memory utilization. The Hadoop in-memory cache employs centralized cache management that can cache both input and intermediate data. In this case, the NameNode is responsible for coordinating all the DataNode off-heap caches in the cluster. The NameNode periodically receives a cache report from each DataNode, describing all of the data blocks cached on a given DataNode. The NameNode manages DataNode caches by piggybacking cache and uncached commands on the DataNode heartbeat message. While the OS page cache employs an LRU-like algorithm for its cache replacement, HDFS in-memory caching does not replace previously cached data blocks unless users ask for uncached data blocks. Therefore users must determine manually which data should be cached or uncached which can limit effective usage of the in-memory cache.

Moreover, the LRU policy typically suffers from cache pollution, where unpopular objects can occupy cache space for a long time. For example, in LRU, suppose that a new item is inserted at the top of the cache. If the item is not requested again, it will take some time to be moved down to the bottom of the cache before removing it from the cache. To address this issue, we customize an intelligent cache replacement strategy, SVM-LRU, for the Hadoop environment (H-SVM-LRU). This strategy combines a supervised machine learning algorithm, SVM, with LRU to classify input data into two groups: reused in the future or not. This method determines the victim data that must be uncached (based on their class) to avoid cache pollution.

## 3. Related work

In this section, we first provide an overview of cache replacement algorithms that have been proposed for improving Hadoop performance. We then discuss different intelligent caching mechanisms that use machine learning methods.

## 3.1 Cache replacement policies in Hadoop

In this section, we investigate different cache replacement strategies in Hadoop, including their advantages and disadvantages.

*LIFE* and *LFU-F* are two replacement strategies used in *PacMan* [8] for an in-memory coordinated caching mechanism and data-intensive parallel jobs. In *PacMan*, parallel jobs run numerous tasks concurrently in a wave with the all-or-nothing property. The *LIFE* algorithm evicts data blocks of files with the largest wave-width and results in reducing the average job completion time. *LFU-F* aims to maximize cluster efficiency. For this purpose, it evicts data blocks with less frequent access. Both strategies prioritize incomplete files over completed files for eviction and use a window-based aging mechanism to avoid cache pollution. They first check whether there are data blocks that have not been accessed within a given time window. Among these files, the one with the least number of accesses is chosen.

*Enhanced Data-Aware Cache (EDACHE)* [9] was introduced for caching intermediate results to accelerate MapReduce job execution times. In this strategy, *WSClock* is used as a cache replacement algorithm in which cached items are maintained in a circular list and a clock hand advances around this ring. This algorithm replaces cached items based on their reference bit and the last time used. It first checks the reference bit. If its value is one it means this item is used. The item's reference bit is then reset, its last time used is updated, and the clock hand is advanced. Otherwise, an item with an age greater than a threshold value is evicted. The bottleneck of this mechanism is related to the fact that large blocks lead to long search times for requested contents.

In *collaborative caching*, a *Modified ARC replacement algorithm* [10] was proposed in order to increase the cache hit ratio and improve efficiency. In this strategy, the cache is divided into the recent cache, recent history, frequent cache, and frequent history such that the cache sections contain data blocks and the history sections include references to evicted items. Initially, on a request for a block, the references in the history caches are checked. If present the corresponding block is placed in the recent or frequent cache, otherwise the cache references and serves the request from either of the history caches, which helps in faster caching as well as locating files for initial checks. If references are found in recent history then the block is placed in the recent cache. If the block is found in the recent cache, then it is moved to the frequent cache, hence a hit in either of the history caches removes the reference and places the corresponding block in one of the caches (recent or frequent). Caching a block also involves caching metadata. When either of the caches is fully utilized then a block is evicted from the recent or frequent cache but its reference is placed into its corresponding history. When either of the history caches is fully utilized the references simply drop out of the cache.

An adaptive cache algorithm [11] was designed to cache the partition of tables into the HDFS cache for Big SQL. Selective *LRU-K (SLRU-K)* and *Exponential-Decay (EXD)* are used as online caching algorithms and selective cache insertion to reduce the overhead of inserting items into the HDFS cache. *SLRU-K* takes into account the variable size of the partition and uses a weight heuristic to place the partitions into the cache selectively. It keeps the list of K's last access time for each partition. However, *EXD* maintains only the time of last access to computing the score for each partition that determines the weight of access frequency versus recently used. In *Adaptive SLRU-K and EXD*, the adaptor adjusts its behavior with access patterns of various workloads by automatically adopting the value of their parameters. Maximizing the byte hit ratio and minimizing the byte insertion ratio is the primary aim of the adaptor.

The *block goodness aware cache replacement strategy* [12] was presented in 2017 and uses two metrics for cache management: cache affinity (CA) depends on resources used by the application and block goodness (BG) measures how much a cached data block is worth. For each cached item, this strategy first calculates the BG value based on the data block access count and MapReduce application cache affinity then selects a data block with the lowest BG value for eviction. A data block with the oldest access time will be discarded if there is more than one data block with the same lowest BG value.

The *Cache Affinity Aware Cache Replacement Algorithm* [13] was designed in 2018 and categorizes MapReduce applications based on their cache affinity. This algorithm prioritizes caching input data of applications with high cache affinity. It takes into account the cache affinity of a MapReduce application and data access frequency to calculate the benefit of caching for input data. As a result, it evicts a data block with the lowest caching benefit. If there are some data blocks with identical lowest benefits, it evicts a block based on the LRU policy.

*AutoCache* [14] was developed in 2019 and employs a lightweight gradient-boosted tree (XGBoost) to predict file access patterns on the HDFS cache. In this mechanism, the probability of accessing a file is measured by a probability score which is used as a metric by the cache replacement policy to avoid cache pollution. When the free space of the cache is less than 10%, the eviction operation is started and it continues until the cache capacity becomes lower than 85%. This cache replacement algorithm has a low overhead by limiting computation to a fixed number of files.

In Table 1, we compare these cache replacement strategies in terms of their criteria for eviction and mitigating cache pollution and summarize their advantages and disadvantages.

Table 1: Hadoop cache replacement comparisons

| Replacement strategy | Metrics for eviction | Cache pollution | Advantages | Disadvantages |
|---|---|---|---|---|
| LIFE | Largest wave width and incompleted file | Window age strategy | Reduces average completion time | Effective for short jobs |
| LFU-F | Frequency access and incompleted file | Window age strategy | Maximizes cluster efficiency | Effective for short jobs |
| WSClock | Last time used | No | Decreases execution time | Long search times for large data blocks |
| Modified ARC | Recency and frequency of access | No | Increases cache hit ratio | Needs space for storing history |
| Adaptive cache | The score of each partition | No | Adapts to various workload characteristics | Significant overhead |
| Block goodness aware | Block goodness value and access time | No | Effective for multiple concurrent applications | Needs to calculate block goodness |
| Cache Affinity aware | Cache affinity of application and recency | No | Effective use of cache space | Needs to know the cache affinity of applications |
| AutoCache | probability of accessing the file | Calculating probability score | Reduces average completion time and improves cluster efficiency | File oriented cache |

## 3.2 Intelligent caching mechanisms

In a related application domain, several intelligent caching strategies have been presented that use different machine learning techniques to enhance the performance of web proxy caches. Ali et al. proposed *SVM–LRU, SVM–GDSF*, and *C4.5–GDS* [15] that combined a support vector machine (SVM) and a decision tree (C4.5) with Least-Recently-Used (LRU), Greedy-Dual-Size (GDS), and Greedy-Dual-Size Frequency (GDSF) replacement strategies. In these techniques, web objects are classified into two groups: revisited later or not. These methods use a web proxy log file as a training dataset and different features of web objects such as recency, frequency, size, access latency, and type of object are considered for classification. Experimental results show that SVM-LRU appears to have the best hit ratio.

Employing a Bayesian network, *BN-GDS*, and *BN-LRU* [16] were introduced to improve the performance of cache replacement strategies like Greedy-Dual-Size (GDS), and Least-Recently-Used (LRU). In these strategies, the probability of web objects belonging to the revisited class (and so should be cached) is calculated based on features such as retrieval time, frequency, size, and type. Experimental results suggest that BN-GDS achieves the best hit ratio while BN-LRU has the best byte hit ratio.

Hybrid ELM-LFU [17], a two-level caching scheme for web proxies, was presented in 2018. In the first level, LFU is used for fast caching replacement (due to its low complexity), and thus is suitable for real-time communication. An Extreme Learning Machine (ELM) is used in the second level, applying a single hidden layer feed-forward network where there is no need to adjust the weights. In this mechanism, the chosen web object for eviction in the first level will be placed in the second-level cache. This method features low training times.

In [18], Herotodos et al. designed a framework for moving data automatically through tiered storage in the distributed file system via a set of pluggable policies. For this purpose, they employ incremental learning to find which data should be downgraded or upgraded allowing for adaption to workload changes over time. For downgrading, this method uses different caching replacement strategies like LRU, LFU, Least Recently & Frequently Used (LRFU), LIFE, LFU-F, Exponential Decay (EXD), and XGBoost-based Modeling (XGB). Also, On Single Access (OSA), LRFU, EXD, and XGB are used for the upgrade policy. Experimental results show that XGB is more suitable because it requires minimal storage, has low training overhead, makes useful predictions, and can learn incrementally over time.

PACS-oriented SVM-LRU [19] was proposed for picture archiving and communication systems in 2021. This algorithm calculates the probability of future access to cached items. In this strategy, SVM-LRU has brought some benefits like low training time, low computation, high prediction accuracy, and high hit ratio.

Even though using a caching mechanism has yielded some benefits in the Hadoop environment, some challenges remain. For instance, cache management imposes a heavy load on the NameNode, both in terms of required storage and computational load, potentially degrading performance. Moreover, existing cache replacement policies in Hadoop do not take into account cache pollution and effective use of cache space and they do not apply intelligent caching mechanisms. In this paper, we design a cache replacement mechanism as an approach for overcoming these problems, using SVM to classify data, resulting in improved performance. We choose SVM because its generalization ability can be maximized when training data are scarce, and it can control the misclassification error.

# 4 H-SVM-LRU cache replacement

Support Vector Machine (SVM) [20] is a supervised machine learning technique that is used for binary classification, dividing data into two classes: positive and negative. In this section, we provide H-SVM-LRU implementation details and explain our replacement algorithm that combines LRU with an SVM to classify data into two classes: reuse in the future or not. The aim of this algorithm is to avoid cache pollution and to effectively use cache space, leading to decreased execution times. In this section, we present a framework for intelligent cache replacement based on machine learning and explain the H-SVM-LRU algorithm with an example to illustrate its operation.

## 4.1 The proposed H-SVM-LRU framework

In this section, we propose a framework for an intelligent LRU approach using an SVM and in-memory cache [21], [22], [23] for Hadoop which consists of two functional components: the classification component is responsible for training the data classification by using an SVM classifier and the Hadoop in-memory cache component uses this trained classifier to manage the cache space. Figure 1 gives the system structure and its components which we now explain in detail.

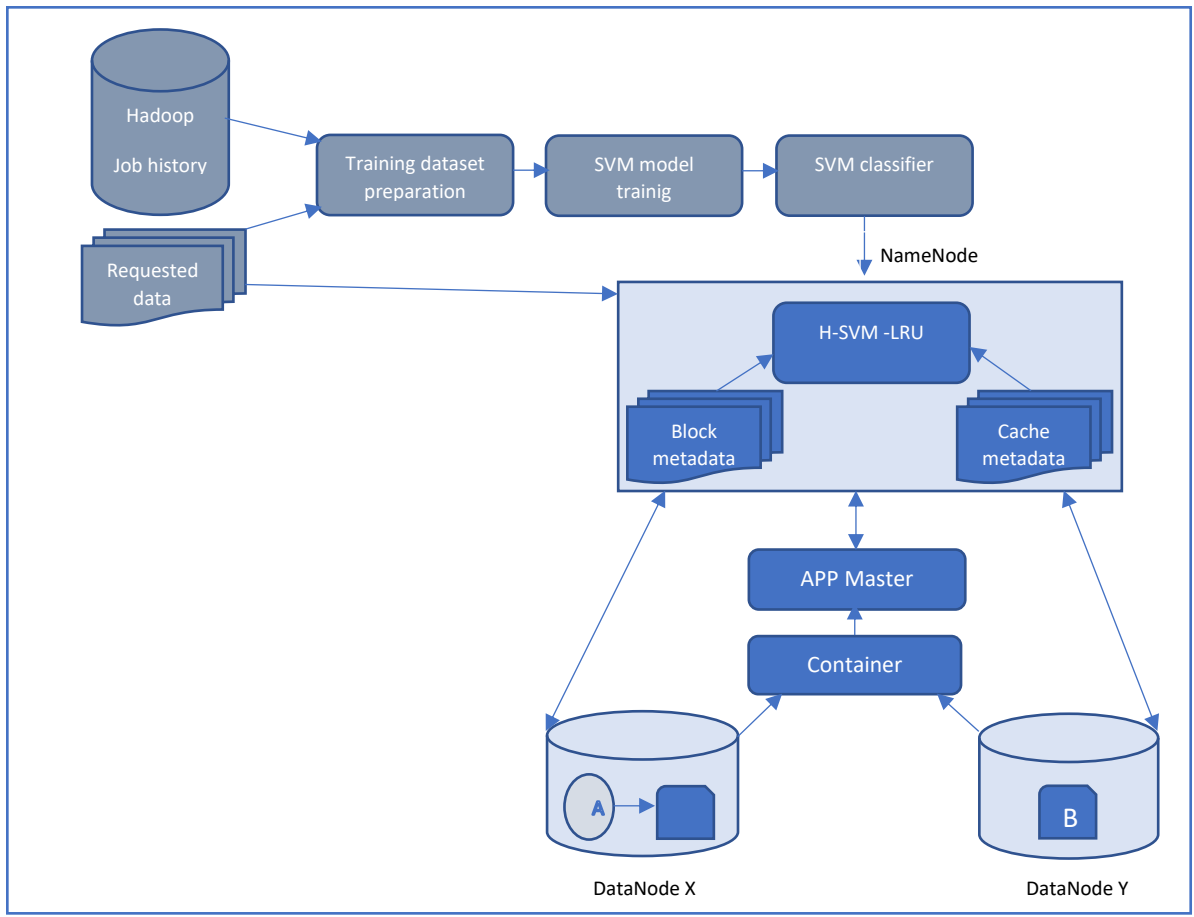

Fig. 1: The proposed H-SVM-LRU intelligent cache replacement strategy framework

The classification component consists of Hadoop job history, training dataset preparation, SVM model training, and SVM classifier. The job history server allows the user to get log information on finished applications. This information can be exploited as a source to extract training data. Next, preprocessing data is employed to normalize data and eliminate outliers. After preparing training data, an SVM classifier is trained. Finally, the classifier is deployed, and SVM-LRU uses this data classification in its cache replacement decision.

The Hadoop in-memory cache component is composed of NameNode, DataNodes, Application Master, and container. The NameNode is responsible for coordinating all the DataNode caches in the cluster and stores two types of metadata: block metadata includes the location of data blocks on DataNodes and cache metadata maps the locations of cached data. The NameNode periodically receives a cache report from each DataNode describing all the blocks cached on a given DataNode. The cache report is used to update cache metadata. For better utilization of the large distributed in-memory caches in Hadoop clusters, each Hadoop container always sends a request to cache the accessed block to the NameNode, and then the NameNode controls which data blocks are added and evicted to and from in-memory caches. We use centralized cache management; as a result, H-SVM-LRU is located on the NameNode. In our system, a container is launched to run a task (either a Map task or a Reduce task) and always sends a request to find cached data blocks. Application Master manages the user job lifecycle and resource needs of individual applications. Each application has a unique, framework-specific Application Master associated with it. It coordinates an application's execution in the cluster and also manages faults.

Assume that a MapReduce application requires two data blocks A and B, where data block A is located on DataNode X and data block B is cached on DataNode Y. The Application Master communicates with the NameNode (which contains the cache metadata) to query the locations of the input blocks and their availability in the cache. A cache miss occurs when looking for data block A and a cache hit occurs for data block B. In the cache miss state, the NameNode looks for block metadata to find the DataNode that contains data block A. Although, there are multiple replicas of a given data block that can be accessed by the query, we choose the first one to reduce search time. We could cache all replicas of this data block in the DataNodes that contain them. In this case, cache replication is identical to data replication and can increase the cache hit rate ratio. In this case, excessive cache space is occupied, conflicting with the goal of our proposed method.

We then use the H-SVM-LRU algorithm and the PutCache(A, X) method to place this data block in the cache. After caching, DataNode X piggybacks the cache report with a heartbeat message and sends it to the NameNode to update the cache metadata. The NameNode informs the Application Master of the location of cached data by using GetCache(A, X), and GetCache(B, Y). When a cache hit occurs, the GetCache method of the H-SVM-LRU algorithm is called to retrieve the cached data. The end result is that not only applications not only do not wait for the data block to get cached but also the probability of accessing cached data has increased via effectively using cache space. It is important to note that applications do not necessarily access all their demand data from the cache, i.e. it is not necessary to wait for data to be cached.

### 4.2 The proposed H-SVM-LRU algorithm

In this section, we describe the proposed H-SVM-LRU algorithm for reducing cache pollution. The ordered dictionary data structure is used to implement the LRU cache because it remembers the order in which keys were first inserted. This data structure removes the first item when the cache does not have sufficient space. When a cache hit occurs and a data block is found in the cache, its key is moved to the end to show that it was recently used. We assume that the victim item is removed from the top of the cache and the recently used items are moved to the bottom of the cache. The proposed H-SVM-LRU algorithm consists of two procedures: GetCache (DataBlock, DataNode) to retrieve data items from the cache, and PutCache (DataBlock, DataNode) to place data blocks into the cache. This algorithm works as follows.

---

**Algorithm 1: H-SVM-LRU on Hadoop**

1- Input R= {$DB_1$, $DB_2$,......,$DB_n$}
2- for each data block $DB_x$ requested by a task $T_i$
3- $DN_y$ ←lookup for $DB_x$ in the cache metadata        //$DB_x$ is cached into $DN_y$
4- if $DB_x$ is in the cache then
5-    begin
6-       cache hit occurs
7-       call GetCache($DB_x$ , $DN_y$)                    //Retrieving $DB_x$ from $DN_y$
8-    else
9-       cache miss occurs
10-      $DN_z$ ←lookup for $DB_x$ in the block metadata  //$DB_x$ is located on $DN_z$
11-      call PutCache($DB_x$, $DN_z$)                     //Place $DB_x$ into the cache
12-   end
13- Procedure GetCache( $DB_x$ , $DN_y$ )
14-    begin
15-       class of $DB_x$ =Apply-SVM (features)
16-       if the class of $DB_x$=1 then                   //$DB_x$ classified as reused class
17-          move $DB_x$ to the bottom of the cache
18-       else                                            //$DB_x$ classified as unused class
19-          move $DB_x$ to the top of the cache
20-    end
21- Procedure PutCache( $DB_x$ , $DN_z$ )
22-    begin
23-       if insufficient space in $DN_z$ cache for $DB_x$
24-          Evict $DB_t$ from top of cache
25-       class of $DB_x$ =Apply-SVM (features)
26-       if the class of $DB_x$=1 then       //$DB_x$ classified as reused class
27-          insert $DB_x$ at the bottom of the cache
28-       else                                 //$DB_x$ classified as unused class
29-          begin
30-             if there are some DB with class unused then
31-                insert $DB_x$ at the end of the unused data list in the cache
32-             else
33-                insert $DB_x$ at the top of the cache
34-          end
35-    end

---

The input is the sequence of data requested by tasks. When the data block $DB_x$ is requested by a task, it searches for the requested data block in the cache metadata to find the data block and the DataNode where it is cached, resulting in a cache hit or cache miss. If data block $DB_x$ is cached at DataNode $DN_y$, the cache hit state occurs and GetCache ($DB_x$, $DN_y$) is called. In this procedure, the SVM predicts whether the class of that data block is that it will be reused in the future or not. The algorithm then moves the data block in the cache based on its class. If the data block is classified by the SVM as an item to be reused, the data block $DB_x$ will move to the bottom of the cache. Otherwise, it will move to the top of the cache to remove it immediately and free cache space.

In the cache miss state, the data block does not exist in the cache, so it can be cached for future uses. For this purpose, the block metadata is first used to find the location of the requested data block on a DataNode (for instance, $DN_z$), and a request to cache this data block is sent by calling PutCache ($DB_x$, $DN_z$). This method first checks the cache capacity as to whether it has sufficient space or not. If there is insufficient space, then it evicts the top item from the cache and the SVM predicts the class of the new item. If the data block is classified to be reused then it is placed at the bottom of the cache; otherwise, it is placed at the end of the unused data list, located at the top of the cache.

H-SVM-LRU can efficiently remove unwanted items at an early stage to make space for new data blocks. By using this mechanism, cache pollution can be reduced, and the available cache space can be utilized more effectively. If all data blocks in the cache have the same class, the proposed algorithm is identical to LRU and only considers the recently used metric for data eviction. Algorithm 1 presents H-SVM-LRU as a cache replacement strategy for the Hadoop environment.

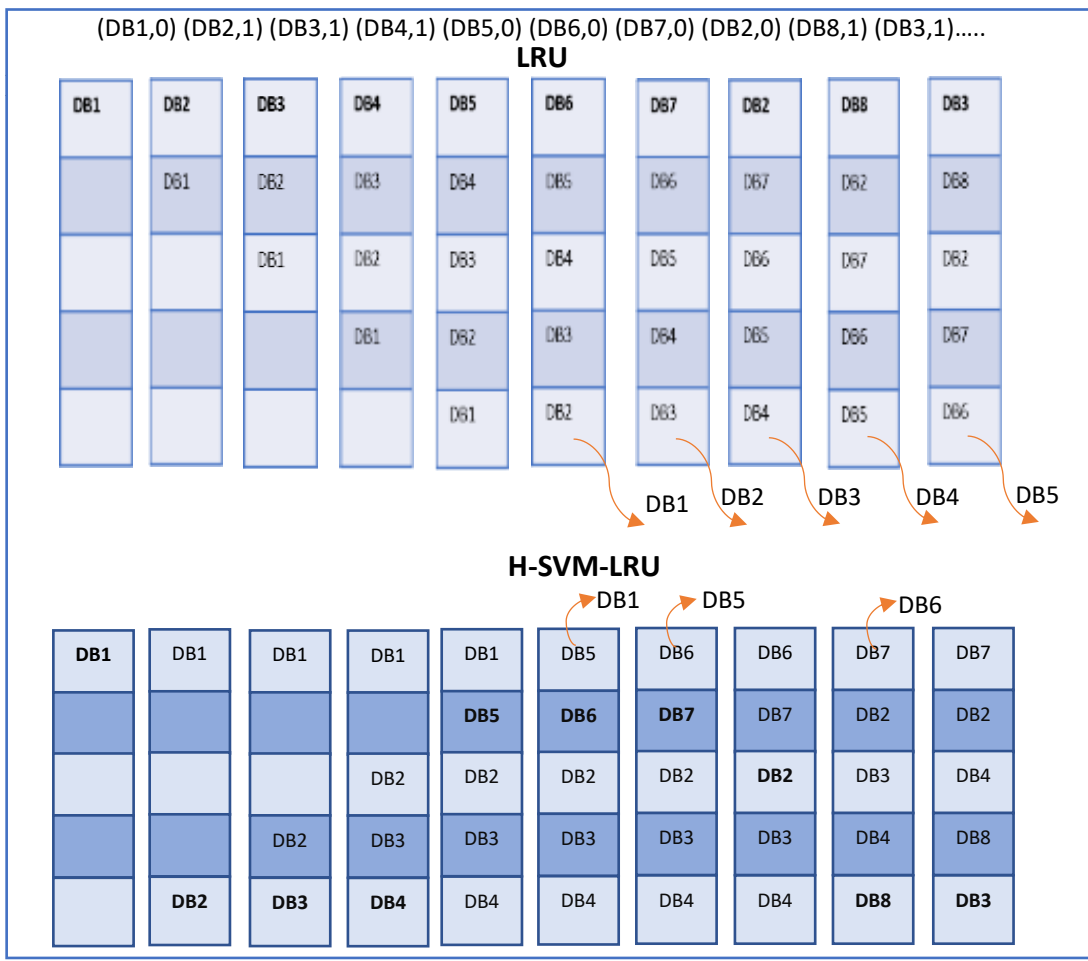

Fig. 2: Example LRU and H-SVM-LRU replacement mechanisms

In order to understand the benefits of the proposed intelligent LRU, we provide an example to compare LRU with H-SVM-LRU. Figure 2 illustrates an example of these two algorithms and compares these two methods. We assume the cache capacity is capable of storing up to five data blocks of the same size. We consider the following subset of the sequence of data blocks with their associated class: (DB1,0) (DB2,1) (DB3,1) (DB4,1) (DB5,0) (DB6,0) (DB7,0) (DB2,0) (DB8,1) (DB3,1). It can be observed that data blocks DB1, DB5, DB6, and DB7 are not reused in the future, while the other data blocks are reused. In the traditional LRU policy, which does not consider class information, all items are initially stored at the top of the cache, so they need a longer time to move down to the bottom of the cache where the least recently used items are stored for evicting. By contrast, the proposed intelligent LRU considers the class of data blocks to determine the victim item that should be evicted. This strategy stores the reused data blocks at the bottom of the cache (DB2, DB3, DB4, and DB8). On the other hand, if data blocks are properly classified as not reused, then they will be stored at the top of the cache (DB1, DB5, DB6, and DB7). Therefore, the unused data blocks are removed earlier by the H-SVM-LRU mechanism to make space for new data blocks.

It can be observed that in the LRU policy, the data blocks DB2 and DB3 are evicted although they are reused in the near future, resulting in cache misses. In H-SVM-LRU, these data blocks have not been evicted leading to an increase in the cache hit ratio. It can be noted that the proposed intelligent LRU approach efficiently removes unused data blocks early to make space for new data blocks. Therefore, cache pollution is decreased, and the available cache space is exploited efficiently. Moreover, the hit ratio and byte hit ratio can be improved. This aspect is discussed in more detail in Section 6.

## 5. H-SVM-LRU implementation

In this section, we provide details for the two phases of H-SVM-LRU implementation, training data preparation, and model training.

### 5.1 Training data preparation phase

This phase consists of four steps: data collection, feature selection, providing target labels, and data preprocessing. We now explain each step in more detail:

- *Data collection*: We consider two independent scenarios: request awareness and non-request awareness.

    In the first scenario, the sequence of requested data is determined by the tasks, and we consider the following data features: size, recency, frequency, and type (input data of Map tasks, intermediate data, and output of Reduce tasks). Table 2 describes these features.

    In the second scenario, we use the ALOJA [24] Hadoop dataset that gathers training data by executing various workloads from the Intel Hi-Bench benchmark suite [25], [26] a comprehensive benchmark suite for Hadoop consisting of a set of Hadoop programs, including both synthetic micro-benchmarks and real-world Hadoop applications. Then we extract some useful features from the Hadoop job history [27], consisting of log information of MapReduce jobs.

In the request awareness scenario, the demand data are predefined. In other words, training data have a label, therefore target labels do not need to be generated. This allow us to consider fewer features than in the second scenario where the requirement to generate target labels may require the consideration of more features.

Table 2: Features for the request-awareness scenario

| Feature name | Description |
| --- | --- |
| Type | The input of the Map task, intermediate data, and the output of Reduce task |
| Size | Size of data blocks in MB |
| Recency | Time last used |
| Frequency | Number of uses |

- *Feature selection*: It is important to choose suitable features for ensuring model performance while reducing overfitting and computational demand. There are different metrics to select features
1. Selecting features based on missing values
2. Selecting features based on variance
3. Selecting features based on correlation with other features
4. Selecting features based on model performance

Table 3: Features for the non-request-awareness scenario to provide target label

| Feature name | Feature type | Description |
| --- | --- | --- |
| JobName | Job | Name of job: WordCount, Sort, Grep, Sort, etc |
| MapsTotal | Job | The total number of Map tasks |
| MapsCompleted | Job | The number of completed Map Tasks |
| ReducesTotal | Job | The total number of Reduce tasks |
| ReducesCompleted | Job | The number of completed Reduce tasks |
| Job-Status | Job | Valid values of job state are: New, Initiated, Running, Succeeded, Failed, Killed, and Error |
| Cache Affinity | Job | Cache affinity of application: Low, High, Medium |
| Start time | Job | The time the job started (in ms) |
| Finish time | Job | The time the job finished (in ms) |
| Task type | Task | Map or Reduce task |
| Task status | Task | The states of the task are: New, Scheduled, Running, Succeeded, Failed, and Killed |
| AvgMapTime | Task | The average time of a Map task (in ms) |
| AvgReduceTime | Task | The average time of a Reduce task (in ms) |
| Progress | Task | The progress of the task as a percentage |

In this step, we select the features given in Table 3. For simplicity, we ignore the size of the data and recently used data features because input data are split into data blocks of the same size, and the recently used data feature is taken into account by the LRU policy.

- *Providing target labels:* Since the training dataset does not have target labels, we should provide them. For this purpose, we use a scenario that is based on job status and the status of its Map tasks and Reduce tasks to provide a label for requested data of tasks. Table 4 describes different cases for this scenario.
- *Dataset preprocessing*: The last step of preparation of the training dataset is data preprocessing which includes the elimination of irrelevant data, unnecessary fields, and data normalization.

Table 4: Guidelines to provide target labels

| Job status | Map task status | Reduce task status | Input Map task label | Input Reduce task label | Rationale |
| --- | --- | --- | --- | --- | --- |
| New | New | New | Not reused | Not reused | The job is waiting in a queue. |
| Initiated | Scheduling | Waiting | Reused | Not reused | The outputs of the Map tasks have not been generated yet. |
| Running | Running | Waiting | Reused | Not reused | The outputs of the Map tasks have not been generated yet. |
| Running | Succeeded | Scheduling | Not reused | Reused | If the input of Reduce is the output of the completed Map task. |
| Running | Succeeded | Running | Not reused | Reused | The map task has been completed. |
| Running | Failed | Waiting | Not reused | Not reused | The Map task is failed and cannot generate intermediate data. |
| Running | Succeeded | Failed | Not reused | Not reused | Map task is completed and Reduce task is failed and cannot continue. |
| Running | Killed | Waiting | Reused | Not reused | The killed task may execute on another node (speculative task) |
| Running | Succeeded | Killed | Not reused | Reused | The failed task may execute on another node (speculative task) |
| Succeeded | Succeeded | Succeeded | Not reused | Not reused | Job is completed and we do not consider the relationship between jobs and repetitive and recurring jobs |
| Failed | Don't care | Don't care | Not reused | Not reused | Job-status has higher priority than task status |

**5.2 Model training phase**

In this phase, we use the Scikit-Learn library in Python to implement an SVM for classifying data. The training process includes two steps: choosing the best kernel function and evaluating the trained model.

- *Choosing the best kernel function:* SVM has several available kernel functions that can be used for training, including polynomial, sigmoid, linear, and RBF. We evaluate the performance of kernel functions by using the confusion matrix to choose an appropriate kernel function for the training dataset. A confusion matrix is a table that is often used to describe the performance of a classification model. The metrics that are used in the confusion matrix method for investigating the correctness of classification are:
- *Recall*: The ability of a classification model to identify all relevant instances.
- *Precision***:** The ability of a classification model to return only relevant instances.
- *F1 score***:** This metric combines recall and precision using the harmonic mean.

The formulas for calculating these metrics are as follows:

$$\text{Recall} = \frac{TP}{TP+FN} \qquad \text{Precision} = \frac{TP}{TP+FP} \qquad \text{F1 Score} = 2 * \frac{Precision * Recall}{Precision + Recall}$$

We choose the RBF function as a kernel function for our dataset because it demonstrated the best performance. The experimental results are reported in Table 5.

Table 5: Evaluation of different kernel functions

| Kernel function | Precision | | Recall | | F1-score | | Accuracy |
|---|---|---|---|---|---|---|---|
| Linear | 0 | 0.67 | 0 | 1 | 0 | 0.8 | 0.71 |
|  | 1 | 1 | 1 | 0.33 | 1 | 0.5 |  |
| RBF | 0 | 0.8 | 0 | 1 | 0 | 0.81 | 0.85 |
|  | 1 | 0.65 | 1 | 0.7 | 1 | 0.75 |  |
| Sigmoid | 0 | 0.57 | 0 | 1 | 0 | 0.73 | 0.57 |
|  | 1 | 0 | 1 | o | 1 | 0 |  |

- *Evaluating the trained model*: The dataset is divided randomly into training data (75%) and testing data (25%). In this phase, we use testing data and cross-validation methods to evaluate the training model and its prediction accuracy. The resulting prediction accuracy is 83%. In other words, the probability of misclassification is low. However, if misclassification occurs in reused data it is possible to evict them before reuse which leads to increased cache misses. In contrast, if misclassification occurs for unused data cache pollution may result.

## 6. H-SVM-LRU evaluation

In this section, we explain the experimental environment including software and hardware configurations, and set some Hadoop configuration parameters. Our evaluation is divided into two sections: the H-SVM-LRU performance evaluation and the investigation of the impact of the proposed algorithm on Hadoop performance. We first evaluate the efficiency of the proposed algorithm by using the cache hit ratio as the performance metric. Finally, we perform experiments to present the impact of the H-SVM-LRU cache replacement policy on overall Hadoop performance.

### 6.1 Experimental setup

For our experiments, we use a cluster consisting of a single NameNode and nine DataNodes located in the same rack.

- *Hardware configuration*: These nodes are connected via a 10 Gigabit Ethernet switch. Each node is configured with an Intel Core i7-6700 processor, 16 GB memory, and a one TB hard disk.
- *Software configuration*: We use Ubuntu14.04 as an operating system and JDK 1.8, Hadoop version 2.7 (which employs in-memory caching), and Intel HiBench version 7.1.
- *Hadoop configuration parameters:* The block size of files in HDFS is chosen to be one of two values, 64 MB or 128 MB, the number of cache replicas is set to one, and data replication is set to 3. The memory size for Map task, Reduce task, and node manager is 1GB, 2GB, and 8 GB, respectively. The maximum size of the cache is set to ١٫٥ GB and we assume that each DataNode in the cluster has the same size cache. Table 5 presents the Hadoop configuration parameters with their values. The remaining Hadoop configuration parameters are set to the default values.
- *MapReduce applications*: As we mentioned earlier, we use Intel HiBench as a Hadoop benchmark suite that contains the following applications: 1) WordCount is a CPU-intensive application that frequently occurs for each word in a text file. 2) Sort is a typical I/O-bound application that sorts input data. 3) Grep is a mix of CPU-bound and I/O-bound operations that searches for a substring in a text file. These three applications are supported by Hadoop. 4) Join is a multiple-stage application such that the results of the previous step are used as input for the next step. 5) Aggregation (supported by Hive) is used for the aggregation operation in a query.
- *Dataset:* For carrying out experiments, we use the Gutenberg dataset [28] as input data for the WordCount application to evaluate its execution time based on different input data. As we mentioned earlier in the implementation section, the ALOJA dataset is used as a training dataset for the SVM model. The applications use input files generated by a random text generator.

Table 6: Hadoop parameters

| Hadoop property name | Hadoop property value |
|---|---|
| Dfs.replication | 3 |
| Dfs.blocksize | 64M or 128M |
| Mapreduce.map.memory.mb | 1024 |
| Mapreduce.reduce.memory.mb | 2048 |
| MapReduce.jobhistory.webapp.address | Master:19888 |
| MapReduce.reduce.speculative | False |
| MapReduce.map.speculative | False |
| Mapred.map.tasks.speculative.execution | False |
| Mapred.reduce.tasks.speculative.execution | False |

**6.2 Metrics**

In these experiments, we consider three key metrics to evaluate our proposed algorithm: The first, cache hit ratio is used to evaluate the performance of the proposed H-SVM-LRU cache replacement policy. The other two metrics, job execution time and normalized run time are used for determining the impact on Hadoop performance. In the following, we explain these three metrics:
- *Hit ratio and byte hit ratio:* These are two major factors to evaluate the performance of the cache replacement strategy. Hit ratio relates the number of cache hits to the total number of requests and byte hit ratio relates the number of bytes obtained from the cache to the total number of bytes requested. It is very difficult for a cache replacement strategy to simultaneously optimize these two metrics because improving the hit ratio usually favors small-sized items over large-sized items, leading to a reduced byte hit ratio. In contrast, strategies that tend to increase the byte-hit ratio and prefer large-sized items typically decrease the hit ratio. In the experiments, we only consider the hit ratio because data blocks have the same size.

- *Job execution time*: This plays a vital role in Hadoop performance improvement, and it is related to data access time. The data access time decreases significantly if we can access data from the cache instead of the disk, reducing the job execution. To calculate the average job execution time, we run each application five times.
- *Normalized run time*: For each application in a workload, its run time is normalized based on the Hadoop original (Hadoop no-cache). The average normalized time for applications in one is then calculated to evaluate overall Hadoop performance [11], [12], [27].

**6.3 H-SVM-LRU performance**

For carrying out experiments to calculate the cache hit rates, we consider two data block sizes: 64MB, and 128 MB. The input data size is 2GB with the same sequence of requested data for each mechanism and the cache size is the same in all DataNodes (1.5 GB). We then calculate the cache capacity based on the maximum number of data blocks that can be cached, which was varied between 6-12 for 128 MB block size and 6-24 for 64 MB block size. Figure 3 presents cache hit ratio graphs for block sizes 64 MB and 128 MB.

In Figure 3, we can observe that by increasing the cache size, the hit ratio is increased for both the LRU and H-SVM-LRU strategies as more requested data can be cached. Also, by increasing the data block size, the cache hit ratio increases, for instance, when the cache size is 6 and the data block size has increased from 64MB to 128MB, the cache hit ratio has approximately doubled because we could cache more data. Both diagrams demonstrate that the hit ratio of H-SVM-LRU is higher than for LRU, in particular when the cache size is small. In order to investigate the performance improvement of H-SVM-LRU over LRU, we calculate the improvement ratio of the performance (IR) based on the hit ratio for each cache size.

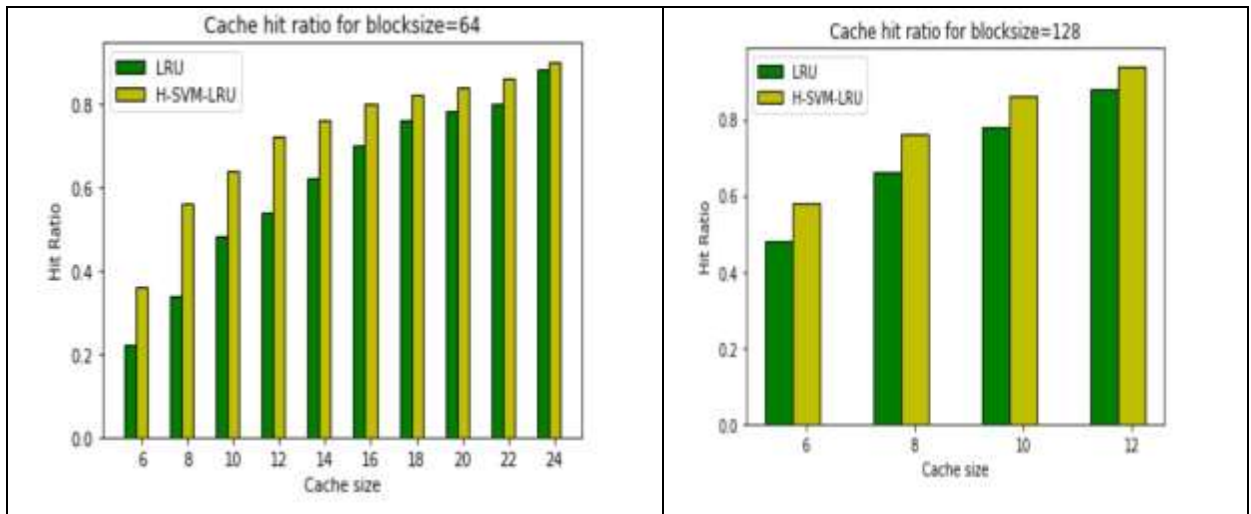

Fig. 3: Cache hit ratio for different cache sizes

Table 7 presents the relative improvement of H-SVM-LRU over LRU for different cache sizes for both 64 MB and 128 MB block sizes. We observe that H-SVM-LRU has the greatest improvement ratio for small cache size and small data blocks, suggesting that H-SVM-LRU is suitable for small cache size because it better avoids cache pollution.

Table 7: Improvement ratio of H-SVM-LRU over LRU based on hit ratio

| Cache size | IR for Data block size (64 MB) | IR for Data block size (128 MB) |
|---|---|---|
| 6 | 63.63% | 20.83% |
| 8 | 64.70% | 15.15% |
| 10 | 33.33% | 10.25% |
| 12 | 33.33% | 6.81% |
| 14 | 22.58% | N/A |
| 16 | 14.28% | N/A |
| 18 | 7.89% | N/A |

**6.4 Impact of H-SVM-LRU on Hadoop performance**

In this section, we carry out two separate experiments to investigate the impact of H-SVM-LRU on Hadoop performance: 1) Job execution time based on different input data sizes. 2) Normalized run time of multiple applications in a workload. For this purpose, we compare Hadoop performance in the following scenarios to extract the impact of the proposed replacement policy on Hadoop performance:

- H-NoCache: The Hadoop original does not utilize HDFS in-memory caching; it is used as a baseline.
- H-LRU: Hadoop uses traditional LRU as a cache replacement policy.
- H-SVM-LRU: H-SVM-LRU is used as a cache replacement policy.

**6.4.1 Job execution time based on different input data sizes**

In this experiment, we consider the job execution time of the WordCount MapReduce application based on different input data sizes for two different data block sizes (64 MB and 128MB). Figure 4 presents job execution time based on input data size for our three scenarios.

As we observe in Figure 4, the job execution time has a significant difference between Hadoop original and Hadoop with cache because, by growing the input data size, the probability of cached data has increased and we could access more data from the cache. When we compare the execution time in H-LRU with H-SVM-LRU, we observe that the execution time of H-SVM-LRU is less than H-LRU because the number of cache hits is greater for H-SVM-LRU. In the second experiment, due to the fact that we use a data block size of 128 MB the difference in execution time between Hadoop original and Hadoop with cache has increased significantly. By increasing the size of the data block we can cache more data than before therefore the byte-hit ratio has increased. In this case, the execution time for H-SVM-LRU is less than for H-LRU because the byte-hit ratio in H-SVM-LRU is more than H-LRU. Therefore, we can conclude that H-SVM-LRU has a lower execution time than the two other scenarios which leads to improved Hadoop performance.

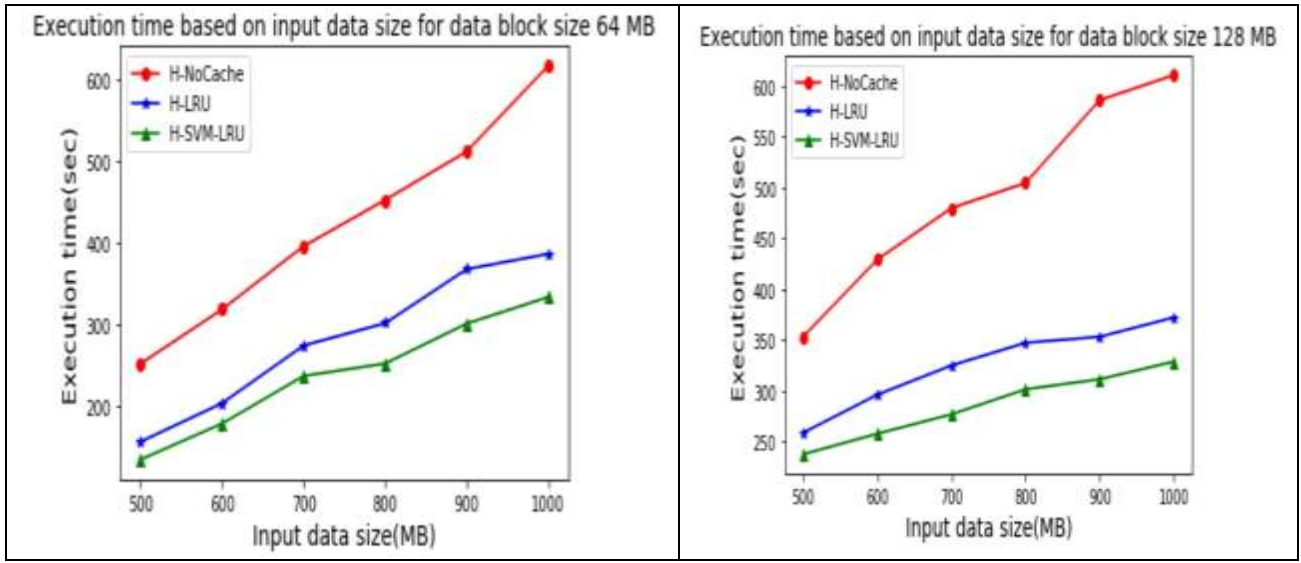
Fig. 4: Job execution time for different input data sizes

**6.4.2 Normalized run time of multiple applications in a workload**

In this experiment, we take into account various workloads which consist of four concurrent MapReduce applications. We assume that all applications in one workload require an equal share of cluster resources. In addition, some applications use the same input data, and data is shared between them, for instance, Grep, WordCount, and Sort use the same input data that are generated by a random text generator, and the data are shared between aggregation and join.

The cache affinity feature [12] determines how to utilize the benefit of cached data in each application such that it can be classified into three categories based on this feature: low cache affinity (Sort), medium cache affinity (WordCount, Join), and high cache affinity (Grep, Aggregation). Therefore, we provide various workloads composed of I/O-bound and CPU-bound applications by considering their cache affinity feature. Table 8 presents the list of workloads with their applications that are used in this experiment.

Table 8: The list of workloads with their applications

| Workload | App1 | App2 | App3 | App4 | Input data size (GB) |
|---|---|---|---|---|---|
| W1 | Aggregation | Grep | Join | WordCount | 257.3 |
| W2 | Aggregation | Grep | Sort | WordCount | 262.9 |
| W3 | Aggregation | WordCount | Grep | Grep | 376.2 |
| W4 | Aggregation | Sort | Grep | Grep | 446.7 |
| W5 | Grep | Grep | Sort | WordCount | 254.3 |
| W6 | Aggregation | Grep | Join | Sort | 377.1 |

In order to compare the Hadoop performance for each workload, we calculate normalized run time based on the Hadoop original (H-No-Cache). Figure 5 illustrates the experimental results. If we compare H-LRU with Hadoop we observe that Hadoop-LRU improves performance by 11.33% and the average improvement of H-SVM-LRU is 16.16%, 4.83% against Hadoop-original and H-LRU respectively. The use of cached data has played a vital role in reducing run time. Also, the number of cache hits in H-SVM-LRU is higher than in Hadoop-LRU as a result of using cache space efficiently. H-LRU and H-SVM-LRU have the best improvement in workload W3 and W5 due to the fact that workload W3 is composed of high cache affinity applications and workload W5 has the most shared data between applications.

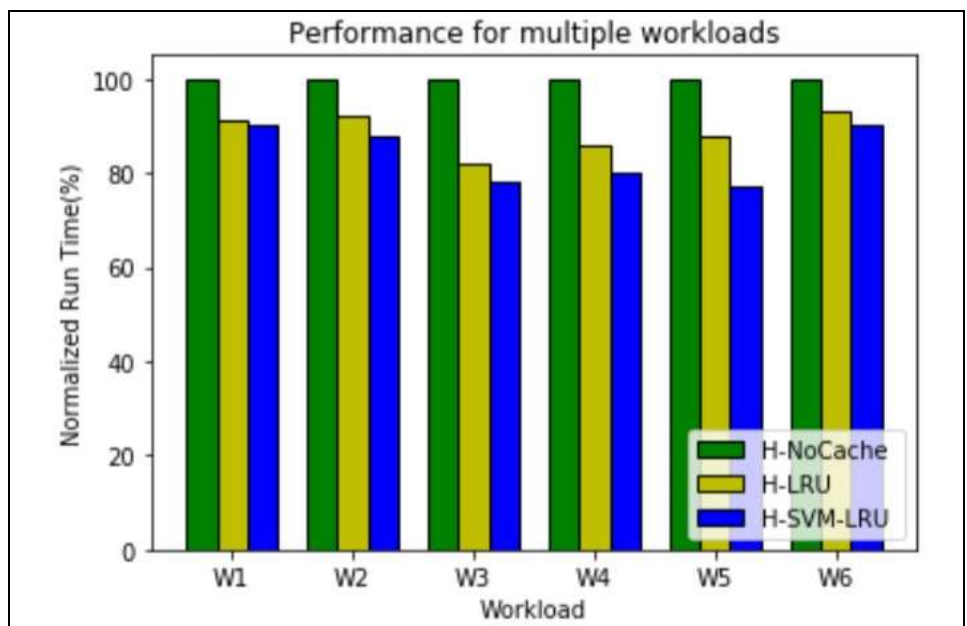

Fig. 5: Normalized run time of different workloads

Figure 6 provides the normalized run times of applications for each workload in the H-SVM-LRU scenario, in order to investigate the impact of H-SVM-LRU on the performance of each application in the workloads. We observe that some I/O-intensive applications like Grep and Sort show significant improvements in their performance because Sort can benefit from reusing cached data by using the same data used by Grep and WordCount, also I/O-bound jobs spend most of their time on reading blocks, therefore they can have increased benefit from cached data. Therefore, the performance of I/O-intensive applications like Sort can be improved when they are combined with other applications with different resource usage patterns. Moreover, multiple-stage applications like Join have difficulty reusing the input files because the output of the previous stage is used as input for the next stage and this usage is not well suited for this caching mechanism.

We conclude that H-SVM-LRU is appropriate for workloads that have a composition of applications with different resource usage and the amount of shared data is high, in other words, this strategy is suitable for jobs that reuse a large amount of data.

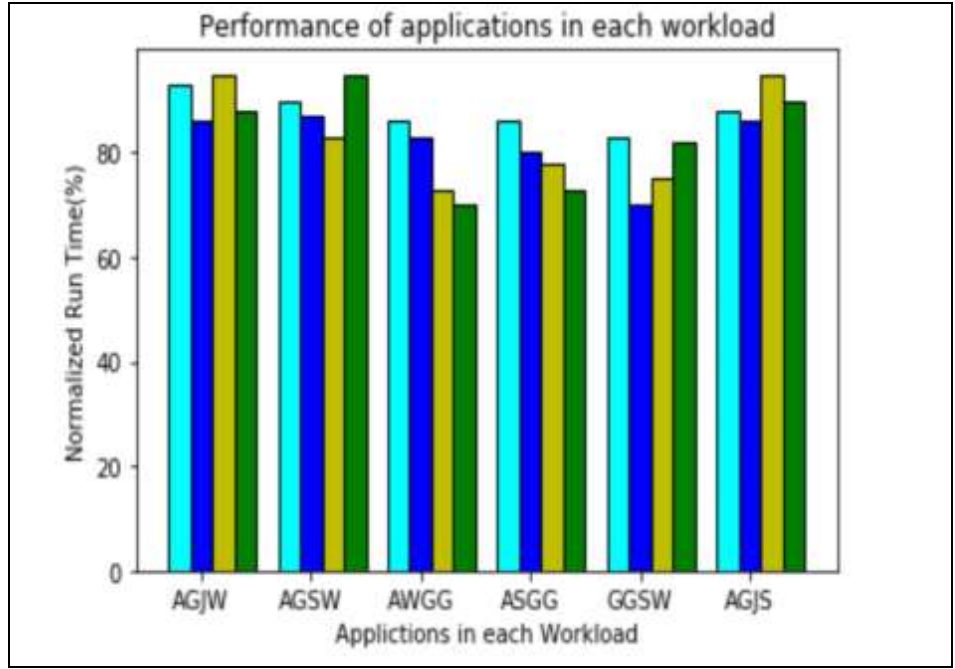

Fig. 6: Normalized run time of applications in each workload

## 7. Conclusion and future work

In this paper, we demonstrated that H-SVM-LRU can efficiently remove unwanted items from the cache at an early stage to make space for new data blocks. By using this mechanism, cache pollution can be reduced, and the available cache space can be utilized more effectively. If all data blocks in the cache have the same class, the proposed algorithm is identical to LRU and only considers the recently used metric for data eviction. Algorithm 1 presents H-SVM-LRU as a cache replacement strategy for the Hadoop environment. We propose H-SVM-LRU as an intelligent cache replacement strategy to improve Hadoop performance. H-SVM-LRU combines SVM with LRU to use the limited cache capacity efficiently and avoids cache pollution by unused data. In this policy, we classify cached data into two groups by using an SVM classifier: data that is reused and unused in the future, and the evicted items are determined based on their class. Experimental results show that the cache hit ratio is improved for H-SVM-LRU via decreasing the frequency of eviction of data that is reused in the future, and we observe the average improvement of H-SVM-LRU is 16.16%, 4.83% against Hadoop-original and H-LRU respectively. This caching mechanism is appropriate for small cache sizes to use its limited space efficiently as well as being suitable for workloads composed of high cache affinity applications with varied resource usage and a large amount of shared data. The advantage of this policy includes increasing the number of cache hits which decreases data access time. This is turn reduces the job execution time resulting in a positive impact on overall Hadoop performance. While the training time is a limitation of this approach, this is somewhat mitigated by the training time being independent of the execution time. The major limitation of this study is that the lack of a labeled training dataset required extra computational overhead, particularly when considering non-request awareness scenarios. Our future plans include examining H-SVM-LRU on a large cluster to evaluate its scalability and extend intelligent caching by applying machine learning techniques to prefetch requested data from HDFS.